\documentclass[intlimits,twoside,a4paper]{article}

\usepackage{amsmath,amssymb}
\usepackage{graphicx}

\usepackage[T2A]{fontenc}
\usepackage[cp1251]{inputenc}

\usepackage[eqsecnum]{cmpj2}

\newcommand{\non}{\nonumber \\}
\newcommand{\be}{\begin{equation}}
\newcommand{\ee}{\end{equation}}
\newcommand{\bea}{\begin{eqnarray}}
\newcommand{\eea}{\end{eqnarray}}
\newcommand{\lp}{\left (}
\newcommand{\rp}{\right )}
\newcommand{\lbr}{\left [}
\newcommand{\rbr}{\right ]}

\newcommand{\ld}{\left .}
\newcommand{\rdo}{\right .}
\newcommand{\ve}[1]{{\bf #1}}
\newcommand{\vk}{\ve{k}}
\newcommand{\rhok}{\rho_{\ve{k}}}
\newcommand{\rhomk}{\rho_{-\ve{k}}}
\newcommand{\cW}{{\cal W}}
\newcommand{\mutb}{\mu_{\tau}+1}
\newcommand{\modt}{|\tau|}
\newcommand{\btPhia}{\beta\tilde{\Phi}(0)}
\newcommand{\brho}{\bar{\rho}}
\newcommand{\avg}[1]{\left \langle #1 \right \rangle}
\newcommand{\Delb}{\Delta_1}

\issue{2013}{16}{2}{23007}

\doinumber{10.5488/CMP.16.23007}

\title[Order parameter of a three-dimensional Ising-like system]%
{Order parameter of a three-dimensional Ising-like system in
the simplest and higher non-Gaussian approximations}
\author[I.V. Pylyuk]{I.V. Pylyuk\thanks{E-mail: piv@icmp.lviv.ua}}
\address{Institute for Condensed Matter Physics of the National Academy
of Sciences of Ukraine, \\
1 Svientsitskii St., 79011 Lviv, Ukraine}

\date{Received February 24, 2013, in final form April 10, 2013}%

\authorcopyright{I.V. Pylyuk, 2013}

\begin{document}

\maketitle

\begin{abstract}
The application of the collective variables method to the study of
the behaviour of nonuniversal characteristics of the system in
the critical region is illustrated by an example of the order
parameter. Explicit expressions for the order parameter
(the average spin moment) of a three-dimensional uniaxial magnet are
obtained in approximations of quartic and sextic non-Gaussian
fluctuation distributions (the $\rho^4$ and $\rho^6$ models,
respectively), taking into account confluent corrections.
Some distinctive features appearing in the process of calculating
the order parameter on the basis of two successive non-Gaussian
approximations are indicated. The dependence of the average spin moment
of an Ising-like system on the temperature and microscopic parameters
is studied.
\keywords three-dimensional Ising-like system, critical behaviour,
non-Gaussian measure density, order parameter
\pacs 05.50.+q, 64.60.F-, 75.10.Hk
\end{abstract}

\section{Introduction}

The author is pleased and honored to contribute to the festschrift dedicated
to the 60th anniversary of Professor Mykhailo Kozlovskii. Mykhailo
Kozlovskii is an excellent scientist in the field of theoretical physics,
one of the founders of a new approach to the description of phase
transitions in statistical physics. This approach is beneficially used
in the present publication.

The research is devoted to the theory of phase transitions and critical
phenomena, which are the subjects of wide-range studies (see,
for example, \cite{bmo190,dmo196,kmo100,plmmo106,nomo110}).
Close to a continuous phase transition, fluctuations on all length
scales make contributions and render the perturbative approach
problematic~\cite{dc105}. Therefore, one should resort
to the non-perturbative methods \cite{krm111} such as the exact and
rigorous analytic solution methods (transfer matrix methods,
combinatorial methods, Bethe-ansatz), conformal field theory analysis,
non-perturbative renormalization-group (RG) analysis, numerical
transfer-matrix calculations, Monte Carlo (MC) simulations. The exact
solutions \cite{rev744,mwmo173,bmo189} and the conformal field
analysis \cite{fmsmo197} are powerful tools to determine the critical
exponents of two-dimensional models. However, these approaches are not
helpful in most of the three-dimensional (3D) cases. The MC method can
be used here. At present, we see a growing interest in the application
of MC-based non-perturbative approaches and the development of powerful
MC simulation tools for 3D systems \cite{pvp109,mmm111}.
An advantage of the MC method is its capability of performing smart moves
that can be tuned to each physical system or phenomenon.
The MC method allows one to reach any desired accuracy.
An alternative method, which has been successful in the study of phase
transitions, particularly in two-dimensional systems, is that of
numerical transfer-matrix calculations combined with finite-size-scaling
analysis \cite{rnr193}. This method is non-perturbative and does not require
foreknowledge of special temperatures (such as the critical temperature)
or of critical exponents. Two main formulations of the non-perturbative
RG [the Wilson approach (also called Wilson-Polchinski) and
the ``effective average action'' approach] are discussed in resent
publications \cite{bb101,btw102,d107,c112}. The exact non-perturbative
RG equations are known for various models on fractal lattices (see,
for example, \cite{rm195} and references therein). Moreover, a rigorous
RG analysis has been made in four dimensions \cite{ht187}. In most other cases, the non-perturbative RG equations cannot be solved exactly,
and suitable truncations must be used,  such as
the derivative expansion \cite{ams196,ams298,mt199,mt201,cdm203}
(limited to the small momentum behaviour) or its adaptation to
arbitrary values of the momenta \cite{bmw106,bbc112}.

In this paper, the mathematical description is performed within
the method of collective variables (CV) \cite{ymo287,ykpmo101}, which
is similar to the Wilson non-perturbative RG approach (integration
on fast modes and construction of an effective theory for slow
modes) \cite{rev14674}.

The object of investigation is a 3D Ising-like
system with an exponentially decreasing interaction
potential $\Phi(r_{\ve{i}\ve{j}})=A\exp(-r_{\ve{i}\ve{j}}/b)$, where
$A$ is a  constant, $b$ is the radius of effective interaction, and
$r_{\ve{i}\ve{j}}$ is the distance between spins located at the sites
$\ve{i}$ and $\ve{j}$ of a simple cubic lattice with period $c$ and
$N$ sites (see, for example, \cite{ykpmo101,pk201,ypk102}). The Ising
model, which is simple and convenient for mathematical analysis, is
widely used in the theory of phase transitions for the analysis of
the properties of various magnetic and nonmagnetic systems (ferromagnets,
antiferromagnets, ferroelectrics, binary mixtures, lattice model of
liquids, etc.). The critical behaviour of the 3D Ising universality
class was also discovered in systems with strong and electroweak
interactions observed in high-energy physics.

The CV method allows one to calculate a partition function of the system
and to obtain universal quantities (critical exponents) and nonuniversal
characteristics (the phase-transition temperature $T_\mathrm{c}$ and
thermodynamic functions near $T_\mathrm{c}$) by using a unified
approach \cite{ymo287,ykpmo101,ypk102}. The available methods
make it possible to calculate universal quantities to a quite high
degree of accuracy (see, for example, \cite{lf189,l494,gz298,pv102}).
An advantage of the CV method lies in the possibility to obtain and
analyse the expressions for thermodynamic characteristics as functions of
microscopic parameters of the system (i.e., the lattice constant and
parameters of the interaction potential). The term collective variables
is applied to a special class of variables specific to each individual
physical system. The set of CV contains variables associated with the order
parameters. For this reason, the phase space of CV is most natural in
describing a phase transition. For magnetic systems, the CV $\rhok$ are
the variables associated with the modes of spin-moment density oscillations,
while the order parameter is related to the variable $\rho_0$,
in which the subscript ``0'' corresponds to the peak of the Fourier
transform of the interaction potential. The CV method is based on
the use of non-Gaussian measure densities \cite{ymo287,ykpmo101,rev9789}.
A non-Gaussian density of measure at a zero external field is
represented as an exponential function of the CV, the argument of which
contains, along with the quadratic term, higher even powers of
the variable with the corresponding coupling constants. The simplest
non-Gaussian measure density is the quartic one (the $\rho^4$ model)
with the second and the fourth po\-wers of the variable in the exponent.
It is followed by the higher sextic measure density (the $\rho^6$ model)
whose exponent includes the sixth power of the variable in addition to
the second and the fourth powers, etc. The starting point of the problem
statement in the CV method is the Hamiltonian of the system. After
passing to the CV set, the Jacobian of the transition from the spin
variables to the CV is calculated to obtain a partition function
functional of the Ginzburg-Landau-Wilson type. The partition
function of the spin system is integrated over the layers
of the CV phase space. The main feature is
the integration of short-wave spin-density oscillation
modes, which is generally done without using perturbation theory.
For this purpose, we divide the phase space of the CV $\rhok$ into
layers with the division parameter $s$. In each $n$th layer
(corresponding to the region of wave vectors
$B_{n+1} < k \leqslant  B_n$, $B_{n+1} = B_n/s$,
$s > 1$), the Fourier transform of the potential is
replaced by its average value (the arithmetic mean in the given case).
To simplify the presentation, we assume that the correction for the
potential averaging is zero, although it can be taken into account if
necessary \cite{ymo287,ykpmo101}. The inclusion of this correction leads to
a nonzero value of the critical exponent $\eta$ characterizing
the behaviour of the pair-correlation function at the critical
temperature $T_\mathrm{c}$ \cite{ykpmo101,ykp312,pu112}. As a result of
step-by-step calculation of partition function, the number of
integration variables in the expression for this quantity decreases
gradually. The partition function is represented as a product of partial
partition functions $Q_n$ of separate layers and the integral of
the ``smoothed'' effective density of measure $\cW_{2m}^{(n+1)}(\rho)$:
\be
Z=2^N2^{(N_{n+1}-1)/2}Q_0Q_1\cdots Q_n[Q(P_n)]^{N_{n+1}}
\int\cW_{2m}^{(n+1)}(\rho) (\rd\rho)^{N_{n+1}}.
\label{cmp1d1}
\ee
Here, $N_{n+1}=N's^{-3(n+1)}$, $N'=Ns_0^{-3}$, $s_0=B/B'$,
$B'=(b\sqrt 2)^{-1}$, and $B=\pi/c$ is the boundary of
the Brillouin half-zone. The integrand
\be
\cW_{2m}^{(n+1)}(\rho)\!=\!\exp\!\! \lbr -\frac{1}{2}
\sum_{k\leqslant  B_{n+1}}d_{n+1}(k)\rhok\rhomk-
\sum\limits_{l=2}^m\frac{a_{2l}^{(n+1)}}
{(2l)!N_{n+1}^{l-1}}\sum_{k_1,\ldots,k_{2l}\leqslant  B_{n+1}}
\rho_{\vk_1}\cdots\rho_{\vk_{2l}}
\delta_{\vk_1+\cdots+\vk_{2l}} \rbr \!\!
\label{cmp1d2}
\ee
characterizes the corresponding non-Gaussian measure density of
the $(n+1)$th block structure (the $\rho^{2m}$ model). At $m=2$, we have
the quartic density of measure or the $\rho^4$ model. The case of $m=3$
corresponds to the effective sextic measure density or
the $\rho^6$ model. In the exponent of the expression (\ref{cmp1d2}),
the coefficient of the quadratic term
$d_{n+1}(k)=a_2^{(n+1)}-\beta\tilde{\Phi}(k)$ depends on the inverse
thermodynamic temperature $\beta=1/(kT)$ as well as on the Fourier
transform of the potential $\tilde{\Phi}(k)=\tilde{\Phi}(0)(1-2b^2k^2)$,
$\delta_{\vk_1+\cdots+\vk_{2l}}$ is the Kronecker symbol,
$B_{n+1}=B's^{-(n+1)}$. The remaining quantities in expressions
(\ref{cmp1d1}) and (\ref{cmp1d2}) are defined in \cite{ymo287,ykpmo101}.

The approximations of quartic and sextic distributions for modes of
spin-moment density oscillations are used in this paper in studying
the behaviour of the spontaneous order parameter, which exists
in the system at temperatures below the critical value of $T_\mathrm{c}$
(the low-temperature region). The scheme of calculating the order
parameter is described for the two above-mentioned non-Gaussian
approximations. Some distinctions connected with these approximations
are revealed.

\section{Order parameter of the system in the
\texorpdfstring{$\rho^4$}{rho4} model approximation}

The average spin moment is the order parameter of the
system investigated. It is associated with the presence of a nonvanishing value of
the variable $\rho_0$ for $T<T_\mathrm{c}$, at which
there is an extremum of the integrand of the expression for
the long-wave part of the partition function
\be
Z_{\mutb}=\re^{-\beta F'_{\mutb}}\int\exp \lbr \beta\sqrt N\rho_0h+
\tilde B\rho_0^2-\frac{G}{N}\rho_0^4 \rbr \rd\rho_0\,.
\label{cmp2d1}
\ee
The number of the layer $\mu_{\tau}$ in the CV space characterizes
the point of exit of the system from the critical-regime region
at $T<T_\mathrm{c}$ \cite{ykpmo101}. The quantity $h$ is determined by
the value of the constant external magnetic field ${\cal H}$ introduced
in our analysis ($h=\mu_\mathrm{B}{\cal H}$, $\mu_\mathrm{B}$ being
the Bohr magneton). The expression for $-\beta F'_{\mutb}$ corresponding
to the contribution to the free energy of the system from CV $\rhok$
with the values of wave vectors $k\rightarrow 0$ (but not equal to zero)
as well as the temperature-independent coefficients of the expressions
\bea
\tilde B&=&\tilde B^{(0)}\modt^{2\nu}\btPhia\left[1+\tilde B^{(1)}
\modt^{\Delta}+\tilde B^{(2)}\modt^{2\Delta}-\tilde B^{(3)}\modt\right], \non
G&=&G^{(0)}\modt^{\nu}\left[\btPhia\right]^2\left[1+G^{(1)}\modt^{\Delta}+
G^{(2)}\modt^{2\Delta}-G^{(3)}\modt\right]
\label{cmp2d2}
\eea
are given in \cite{ykpmo101,kpy291}. Here, $\nu$ and $\Delta$ are the
critical exponent of the correlation length and the exponent of the
temperature confluent correction (i.e., the correction to scaling),
respectively. The reduced temperature is defined by
$\tau=(T-T_\mathrm{c})/T_\mathrm{c}$. Performing in (\ref{cmp2d1})
the substitution of the variable
\be
\rho_0=\sqrt N\rho,
\label{cmp2d3}
\ee
we obtain
\be
Z_{\mutb}=\re^{-\beta F'_{\mutb}}\sqrt N\int \re^{-NE_0^{(4)}(\rho)} \rd\rho.
\label{cmp2d4}
\ee
Owing to the factor $N$ in the exponent in (\ref{cmp2d4}), the integrand
has a sharp maximum at the point $\brho$ corresponding to
the equilibrium value of the order parameter $\avg{\sigma}$.
In the $\rho^4$ model approximation, the expression
\be
E_0^{(4)}(\rho)=G\rho^4-\tilde B\rho^2-\beta h\rho
\label{cmp2d5}
\ee
defines the fraction of free energy associated with the order
parameter. It corresponds to a microscopic analogue of the Landau free
energy. In contrast to the Landau theory, the temperature dependence of
coefficients in (\ref{cmp2d5}) is nonanalytic [see (\ref{cmp2d2})].

Thus, the evaluation of the order parameter $\avg{\sigma}$ is reduced to
determining the extremum point $\brho$ of the
expression (\ref{cmp2d5}). In the case $h=0$, we find the following form
for $\avg{\sigma}$ within the framework of the $\rho^4$ model:
\be
\avg{\sigma}=\sqrt{\frac{\tilde B}{2G}}=\avg{\sigma}^{(0)}\modt^{\nu/2}
(1+\avg{\sigma}^{(1)}\modt^{\Delta}+\avg{\sigma}^{(2)}\modt^{2\Delta}-
\avg{\sigma}^{(3)}\modt).
\label{cmp2d6}
\ee
All the coefficients in (\ref{cmp2d6}) are functions of microscopic
parameters of the system, i.e., the effective radius of the potential
$b$, the Fourier transform of the potential $\tilde{\Phi}(0)$ for $k=0$
and the lattice constant $c$. The expressions for the quantities
$\avg{\sigma}^{(l)}$ ($l=0,1,2,3$) are presented in \cite{kpy291}.
The values of these quantities for the effective interaction
radius $b=c$ are contained in table~\ref{cmpt1}.
\begin{table}[htbp]
\caption{Coefficients in expressions for the average spin moment
of the system in approximations of the $\rho^4$ and $\rho^6$ models
for some values of the RG parameter $s$.}
\label{cmpt1}
\vspace{1ex}
\begin{center}
\begin{tabular}{|c|c|c|c|c|c|c|}
\hline
$s$ & \multicolumn{4}{|c|}{$\rho^4$ model} &
\multicolumn{2}{|c|}{$\rho^6$ model} \\
\cline {2-7}
& $\avg{\sigma}^{(0)}$ & $\avg{\sigma}^{(1)}$ & $\avg{\sigma}^{(2)}$
& $\avg{\sigma}^{(3)}$ & $\avg{\sigma}^{(0)}$ & $\avg{\sigma}^{(1)}$ \\
\hline\hline
2   & 0.4899 & 1.0809 & 1.0453 & 0.5042 & 0.3747 & 0.7485 \\
2.5 & 0.4384 & 0.8049 & 0.8438 & 0.5041 & 0.3387 & 0.5574 \\
3   & 0.3947 & 0.6528 & 0.6764 & 0.5039 & 0.3107 & 0.4651 \\
\hline
\end{tabular}
\end{center}
\end{table}

For an infinitely weak external field $h$, the average spin moment is
defined by the formula
\be
\avg{\sigma} \approx c_\nu^{-1/2}\avg{\sigma}^{(0)}s^{-(\mutb)/2}
\lp 1+\avg{\sigma}_1\modt^{\Delta}+\avg{\sigma}_2\modt^{2\Delta}-
\frac{1}{2}\modt\rp \lp 1-\frac{t}{3\sqrt 3} \rp,
\label{cmp2d7}
\ee
where $c_\nu=(c_{1k}/f_0)^\nu$. The temperature-independent quantities
$c_{1k}$ and $f_0$ characterize one of the coefficients in the solutions
of recurrence relations for the $\rho^4$ model and one of
the fixed-point coordinates, respectively \cite{ykpmo101}. The quantities
$\avg{\sigma}_1$ and $\avg{\sigma}_2$ appear in
the coefficients $\avg{\sigma}^{(1)}$ and $\avg{\sigma}^{(2)}$ from
(\ref{cmp2d6}). The expression \cite{kpy291}
\be
t=-t^{(0)}\left(1+t^{(1)}\modt^{\Delta}+t^{(2)}\modt^{2\Delta}\right)
\label{cmp2d8}
\ee
allows us to write down the average spin moment in the form
\bea
\avg{\sigma} &\approx& c_\nu^{-1/2}\avg{\sigma}^{(0)}s^{-(\mutb)/2}
\left\{ 1+\frac{t^{(0)}}{3\sqrt 3}+
\left[ \avg{\sigma}_1+\frac{t^{(0)}(\avg{\sigma}_1+t^{(1)})}{3\sqrt 3} \right]
\modt^{\Delta} \rdo \non
& & \ld +\left[ \avg{\sigma}_2+\frac{t^{(0)}(\avg{\sigma}_1t^{(1)}+
\avg{\sigma}_2+t^{(2)})}{3\sqrt 3} \right] \modt^{2\Delta}-
\frac{1}{2} \left[ 1+\frac{t^{(0)}}{3\sqrt 3} \right] \modt \right\}.
\label{cmp2d9}
\eea
For $h=0$, the quantity $t^{(0)}\sim h$ vanishes, and
the relation (\ref{cmp2d9}) goes over into (\ref{cmp2d6}).

\section{Average spin moment of the system in the
\texorpdfstring{$\rho^6$}{rho6} model approximation}

In the previous section, the explicit expression for the average spin
moment of the system is obtained on the basis of the $\rho^4$ model,
taking into account the first and second confluent corrections (which
are determined by the terms proportional to $\modt^{\Delta}$ and
$\modt^{2\Delta}$, respectively). In the present section,
the calculation of the order parameter of the system in the higher
non-Gaussian approximation (the $\rho^6$ model) involves
the first confluent correction.

We shall proceed from the expression coinciding in form with
(\ref{cmp2d4}), where the role of $E_0^{(4)}(\rho)$ is played by
\be
E_0^{(6)}(\rho)=D\rho^6+G\rho^4-\tilde B\rho^2-\beta h\rho.
\label{cmp3d1}
\ee
For the quantities appearing in (\ref{cmp3d1}), we can write
the following relations accurate to within $\modt^{\Delb}$ with known coefficients \cite{ykpmo101}:
\bea
\tilde B&=&\tilde B^{(0)}\modt^{2\nu}\btPhia\left[1+\tilde B^{(1)}
\modt^{\Delb}\right], \non
G&=&G^{(0)}\modt^{\nu}\left[\btPhia\right]^2\left[1+G^{(1)}\modt^{\Delb}\right], \non
D&=&D^{(0)}\left[\btPhia\right]^3\left[1+D^{(1)}\modt^{\Delb}\right].
\label{cmp3d2}
\eea
Here, the exponent of the first confluent correction
$\Delb=-\ln E_2/\ln E_1$ in the case of the $\rho^6$ model is
defined by the eigenvalues ($E_1$ and $E_2$) of the RG linear
transformation matrix (like $\Delta$ for the $\rho^4$ model).
In contrast to the expression (\ref{cmp2d5}) (the $\rho^4$ model),
the formula (\ref{cmp3d1}) contains the additional term, in which
the coefficient $D$ is not equal to zero at $\tau=0$. Consequently,
the expression (\ref{cmp3d1}) is most natural in describing
the limiting case of $T=T_\mathrm{c}$ than the relation (\ref{cmp2d5}).

The value of $\brho$ corresponding to the average spin moment
can be determined from the condition of the extremum
$\partial E_0^{(6)}(\rho)/\partial\rho=0$ or
\be
6D\brho^5+4G\brho^3-2\tilde B\brho-\frac{h}{kT}=0.
\label{cmp3d3}
\ee
It is difficult to obtain the general solution of
equation (\ref{cmp3d3}). However, this equation is easy to solve
in some limiting cases. Let us consider the first case
of this kind when there is no external field. For $h=0$, we obtain
the biquadratic equation
\be
6D\brho^4+4G\brho^2-2\tilde B=0,
\label{cmp3d4}
\ee
in which the substitution of the variable
\be
\brho^2=y
\label{cmp3d5}
\ee
leads to the equation
\be
6Dy^2+4Gy-2\tilde B=0.
\label{cmp3d6}
\ee
Solving this equation and separating temperature explicitly, we arrive at
the following formula for the average spin moment $\avg{\sigma}=
\brho=\sqrt{y}$:
\be
\avg{\sigma}=\avg{\sigma}^{(0)}\modt^{\nu/2}
(1+\avg{\sigma}^{(1)}\modt^{\Delb}).
\label{cmp3d7}
\ee
The expressions for the coefficients $\avg{\sigma}^{(0)}$ and
$\avg{\sigma}^{(1)}$ are given in \cite{ykpmo101,kpd297}.
By analogy with the $\rho^4$ model, these coefficients are not
universal since they depend on microscopic parameters of the system.
The values of $\avg{\sigma}^{(0)}$ and $\avg{\sigma}^{(1)}$ (the case of
the $\rho^6$ model when $b=c$) are presented in table~\ref{cmpt1}.

Let us now consider the second limiting case, $T=T_\mathrm{c}$ or
$\tau=0$, when the solution of equation (\ref{cmp3d3}) can be written
in a simple analytic form. Then, $\tilde B=0$, $G=0$ and we obtain
the equation
\be
6D\brho^5=\frac{h}{kT}\,.
\label{cmp3d8}
\ee
In more compact form, it reads
\be
\brho=\brho_0 h^{1/\delta},
\label{cmp3d9}
\ee
where the critical exponent $\delta$ takes on the value 5. The critical
amplitude $\brho_0$ satisfies the following relation:
\be
\brho_0=\left\{ 6kT_\mathrm{c}s_0^6D_1^{(0)}
\left[\beta_\mathrm{c}\tilde{\Phi}(0)\right]^3 \right\}^{-1/5}.
\label{cmp3d10}
\ee
Here, $\beta_\mathrm{c}=1/(kT_\mathrm{c})$ is the inverse
phase-transition temperature. The quantity $D_1^{(0)}$ appears
in the expression for $D$ from (\ref{cmp3d2}) in
the coefficient $D^{(0)}=s_0^6D_1^{(0)}$. The solution (\ref{cmp3d9})
is obtained naturally. It cannot be found in a like manner within
the framework of the $\rho^4$ model (for more details, see
the next section).

\begin{figure}[ht!]
\centerline{\includegraphics[width=0.53\textwidth]{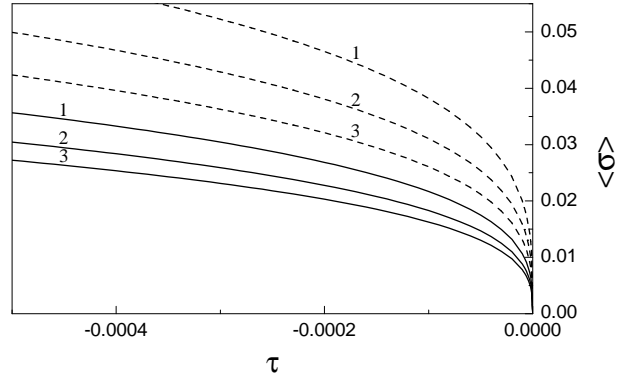}}
\caption{Temperature dependence of the average spin moment of
the system for various values of the RG parameter $s$ within
the framework of the $\rho^4$ model (dashed curves). For comparison,
the average spin moment obtained in the $\rho^6$ model approximation
without involving the confluent correction is also presented
(solid curves). Curves~1, 2, and 3 correspond to $s=2$, $s=2.5$, and
$s=3$, respectively.}
\label{cmpf1}
\end{figure}

\begin{figure}[ht!]
\centerline{\includegraphics[width=0.6\textwidth]{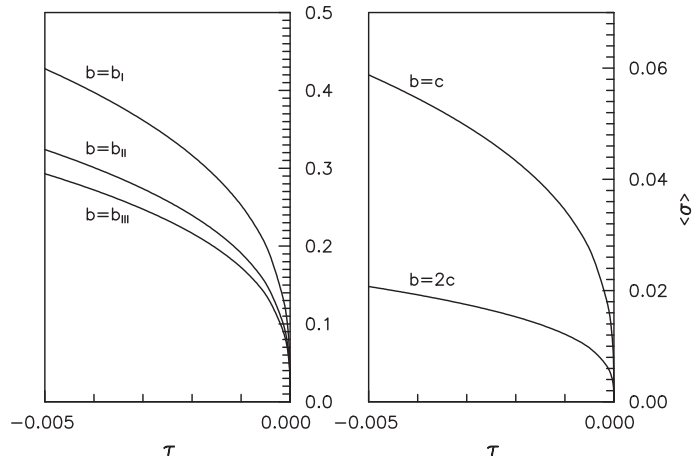}}
\caption{Dependence of the order parameter of the system on
the reduced temperature in the $\rho^6$ model approximation for various
values of the effective radius of the potential $b$:
$b_{I}=c/(2\sqrt 3)$, $b_{\mathrm{II}}=0.3379c$, $b_{\mathrm{III}}=0.3584c$, $c$, and $2c$.}
\label{cmpf2}
\end{figure}

\begin{figure}[ht]
\includegraphics[height=5.3cm]{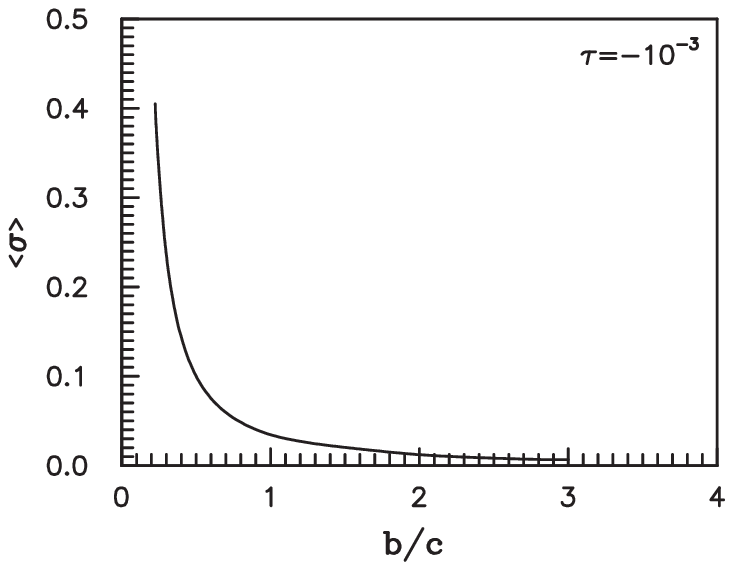}
\hfill
\includegraphics[height=5.3cm]{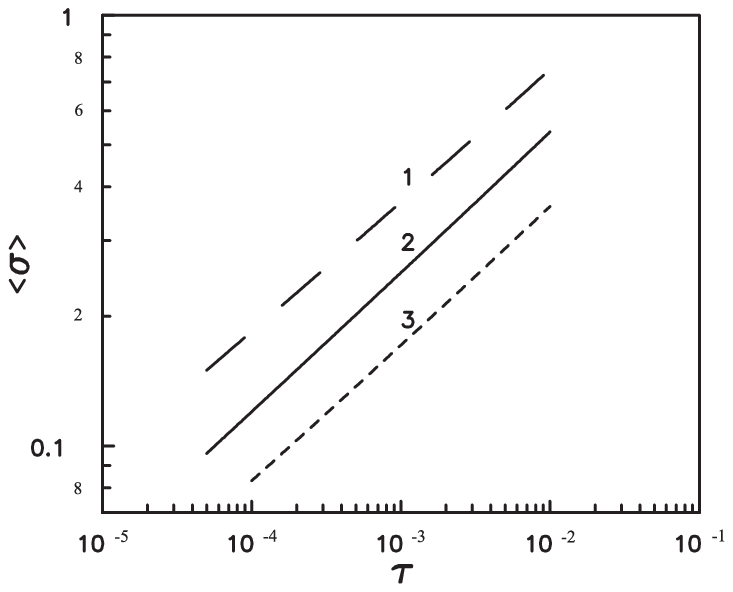}
\\
\parbox[t]{0.47\textwidth}{
\vspace{-0.3cm}
\caption{Behaviour of the average spin moment of the system for
$\tau=-10^{-3}$ with an increasing ratio of the effective radius of
an exponentially decreasing interaction potential to the simple cubic
lattice constant (the $\rho^6$ model).}
\label{cmpf3}
}
\hfill
\parbox[t]{0.47\textwidth}{
\vspace{-0.3cm}
\caption{Order parameter of the 3D Ising model on a simple cubic
lattice. Here, $\tau=|T-T_\mathrm{c}|/T_\mathrm{c}$. Straight line~1
corresponds to the $\rho^4$ model, line~2 corresponds to the $\rho^6$
model, and line~3 corresponds to the results obtained in \cite{lf189}.}
\label{cmpf4}
}
\end{figure}

The explicit expressions (\ref{cmp2d6}) and (\ref{cmp3d7}) make it
possible to study the dependence of the order parameter of the system
in approximations of the $\rho^4$ and $\rho^6$ models on the temperature
and microscopic parameters. The temperature-dependence curves for
$\avg{\sigma}$ near $T_\mathrm{c}$ are shown in figures~\ref{cmpf1} and
\ref{cmpf2} for different values of $s$ and $b$, respectively.
Figure~\ref{cmpf1} demonstrates the results for the effective
interaction radius $b=c$. As is seen from this figure, the dependence
of the average spin moment on the RG parameter $s$ for the $\rho^6$
model is weaker than for the $\rho^4$ model. The $\rho^6$ model
will be the key model in our further study. In figure~\ref{cmpf2} and
below, the plots are obtained for the RG parameter
$s=3$ with allowance for confluent corrections. At $s=3$, the numerical
calculations are performed for $\nu=0.592$, $\Delta=0.475$
(the $\rho^4$ model) \cite{kpy191} and $\nu=0.640$, $\Delb=0.503$
(the $\rho^6$ model) \cite{kpd297}. The value of
$b=b_{\mathrm{I}}=c/(2\sqrt 3)$ corresponds to the interaction between nearest
neighbours, $b=b_{\mathrm{II}}=0.3379c$ corresponds to the interaction between
the nearest and next-nearest neighbours, and $b=b_{\mathrm{III}}=0.3584c$
corresponds to the nearest, next-nearest and third
neighbours \cite{ykpmo101,kpu196}. At these values of $b$ and small
values of the wave vectors $\vk$, the parabolic approximation of
the Fourier transform of an exponentially decreasing interaction
potential corresponds to the analogous approximation of the Fourier
transform for the interaction potentials of the above-mentioned
neighbours. The evolution of the order parameter $\avg{\sigma}$
[see (\ref{cmp3d7}), the case of the $\rho^6$ model] for
$\tau=-10^{-3}$ with an increasing ratio $b/c$ is demonstrated
by the curve in figure~\ref{cmpf3}.

The plots in figures~\ref{cmpf2} and \ref{cmpf3} show that the value of
the order parameter decreases as the potential effective radius $b$
increases. This is the expected result since the condition $b\gg c$
corresponds to the transition to the model with long-range interaction,
which demonstrates a classical critical behaviour.
In the classical case, the order parameter
$\avg{\sigma}_\mathrm{cl}\sim \modt^{\beta_\mathrm{cl}}$
is smaller than the nonclassical order parameter
$\avg{\sigma}\sim \modt^\beta$ ($\beta_\mathrm{cl}=0.5$,
$\beta\approx 0.3$, $\modt\ll 1$). The distribution for
modes of spin-moment density oscillations at $b\gg c$ acquires
the Gaussian form \cite{p599}.

The average spin moment calculated in approximations of the $\rho^4$
and $\rho^6$ models is in accord with the data obtained by
other authors (see figure~\ref{cmpf4}).
The calculations were made for a simple cubic lattice in a zero external
field with the interaction between nearest neighbours. In our
calculations, we put $b=b_{\mathrm{I}}=c/(2\sqrt 3)$. The straight lines~1 and 2
in figure~\ref{cmpf4} are parallel to line~3. This is connected with
the universality of critical exponents. As is clearly seen from
figure~\ref{cmpf4}, the plot for the $\rho^6$ model agrees more closely
with the Liu and Fisher's results \cite{lf189} than the estimates in
the $\rho^4$ model approximation. Liu and Fisher in \cite{lf189}
carried out the numerical analysis of the leading critical amplitudes of
 susceptibility, correlation length, specific heat and spontaneous
magnetization of 3D nearest-neighbour sc (simple cubic), bcc
(body-centered cubic) and fcc (face-centered cubic) Ising models
as well as universal relations between these amplitudes.
Modern estimates of the critical temperature and exponents in
\cite{lf189} are used in conjunction with the biased inhomogeneous
differential approximants to extrapolate the longest available series
expansions to find the critical amplitudes.

\section{Discussion and conclusions}

The main interest of the expressions for thermodynamic functions,
obtained within the CV method, is the direct connection of their
coefficients with the fixed-point coordinates and the microscopic
parameters of the initial Hamiltonian. This shows the capability of
the CV method of being effective in describing both universal
and nonuniversal characteristics of the system as functions of
microscopic parameters. This capability is rather unusual with a RG
approach since, for example, it is well known that the RG of
perturbative field theory is only efficient in the close vicinity of
the critical point and cannot account for a nonuniversal effect of
a specific microscopic parameter of a specific system in a precise way.
In that respect, the CV method is close to the Wilson non-perturbative
RG where the pruning of the field theory (continuum limit) has not yet been
performed. However, the CV method has an advantage over the Wilson
approach, because it starts one step earlier (at the level of an explicit
Hamiltonian) so that the microscopic parameters are directly accessible.

In the present paper, the order parameter (the average spin moment)
of a 3D Ising-like system at $T<T_\mathrm{c}$ is found using the simplest
quartic (the $\rho^4$ model) and higher sextic (the $\rho^6$ model)
non-Gaussian fluctuation distributions. The average spin moment of
the system in these non-Gaussian approximations is determined as
the solution of the corresponding equation. It should be noted that
the solution (\ref{cmp3d7}) does not differ essentially
from the similar one presented above for the case of the $\rho^4$ model
[see (\ref{cmp2d6})]. However, the solution (\ref{cmp3d9}) is obtained
by means of a different route as compared to the $\rho^4$ model.
In the latter case, one solves the cubic equation and excludes
the reduced temperature $\tau$ \cite{ymo287}. For the $\rho^6$ model,
the solution (\ref{cmp3d9}) is simply derived from
equation (\ref{cmp3d3}) by nullifying $\tau$. This demonstrates
that the mentioned solution in the limiting case of $\tau=0$ is
obtained in more natural way for the $\rho^6$ model as compared
to the $\rho^4$ one.

The expressions, presented in this paper in the simplest and higher
non-Gaussian approximations, made it possible to study the behaviour
of the average spin moment as a function of the temperature and
microscopic parameters of the system. It should be emphasized that
the dependence of the order parameter of the system on
the RG parameter $s$ becomes weaker as the form of the non-Gaussian
distribution for modes of spin-moment density oscillations becomes
more complicated (transition from the $\rho^4$ model to the more
complicated $\rho^6$ model). This is confirmed by a direct comparison
of the curves describing the temperature dependence of the order
parameter calculated for the $\rho^4$ and $\rho^6$ models at different
values of the RG parameter $s$ (see figure~\ref{cmpf1}).
The rise in the effective radius of an exponentially decreasing interaction
potential reduces the order parameter of the system
(see figure~\ref{cmpf2}). The variation of the average spin moment
with an increasing ratio of the potential effective radius to the simple
cubic lattice constant is traced in the case of the $\rho^6$ model
(see figure~\ref{cmpf3}). Figures~\ref{cmpf2} and \ref{cmpf3} are
an illustration of the capability of the CV method of controlling the effects
of microscopic parameters of the system.

The plots of the temperature dependence of the order parameter
corresponding to the $\rho^4$ and $\rho^6$ models agree with
the data of other authors (see figure~\ref{cmpf4}). The employment of
the $\rho^6$ model for the investigation of the phase transition by
the CV method gives a more precise definition of the calculation results
and provides the basis for the quantitative analysis of the critical
behaviour of 3D Ising-like systems including the nonuniversal
characteristics \cite{ykpmo101,kpd297}. For this more complicated
$\rho^6$ model, the effect of microscopic parameters on the behaviour
of the average spin moment in the critical region is revealed
in the present study.

%
%

\ukrainianpart

\title{Параметр порядку тривимірної ізингоподібної системи
в найпростішому та вищому негаусових наближеннях}
\author{І.В. Пилюк}
\address{Інститут фізики конденсованих систем НАН України,
вул. Свєнціцького, 1, 79011 Львів, Україна}
%
%
%

\makeukrtitle

\begin{abstract}
\tolerance=3000%
Застосування методу колективних змінних до вивчення поведінки
неуніверсальних характеристик сис\-те\-ми в критичній області проілюстровано
на прикладі параметра порядку. Явні вирази для параметра порядку
(середнього спінового моменту) тривимірного одновісного магнетика
отримано в наближеннях четвірного та шестирного негаусових розподілів
флуктуацій (моделі $\rho^4$ та $\rho^6$ відповідно) з врахуванням
конфлуентних поправок. Вказано на деякі відмінні риси, які проявляються
під час розрахунку параметра порядку на основі двох послідовних
негаусових наближень. Досліджено залежність середнього спінового моменту
ізингоподібної системи від температури та мікроскопічних параметрів.
\keywords тривимірна ізингоподібна система, критична поведінка,
негаусова густина міри, параметр порядку

\end{abstract}

\end{document}